\documentclass[]{spie}  

\usepackage{multirow}%
\usepackage{float}
\usepackage{amsmath,amsfonts,amssymb}
\usepackage{graphicx}
\usepackage[colorlinks=true, allcolors=blue]{hyperref}
\usepackage{tikz}
\usepackage{booktabs}
\usepackage{siunitx}
\usepackage{subcaption}
\usepackage{xspace}
\usepackage[numbers]{natbib}

\title{Design of a three-lens wide field corrector with aspherical surfaces for the 2.34-m VBT}

\author[a]{Nitish Singh}
\author[a]{S. Sriram}
\author[a]{Bharat Kumar Yerra}
\affil[a]{Indian Institute of Astrophysics, II Block, Koramangala, Bengaluru 560 034, INDIA}

\authorinfo{Further author information: Residence: Bhaskara, IIA Guest House, Koramangala II Block, Bengaluru 560034, India\\
Nitish Singh: E-mail: nitish.singh@iiap.res.in, nitiphy@gmail.com\\
Telephone: +91 8077759357}

\begin{document} 
\maketitle

\begin{abstract}
We are developing a compact three-element Wide Field Corrector (WFC) with spherical and aspherical lenses for the 2.34 m Vainu Bappu Telescope (VBT) to enhance its field coverage for imaging and spectroscopic applications. The design consists of three optical elements, with at least one spherical lens movable to serve as an Atmospheric Dispersion Corrector (ADC), while the aspherical elements remain fixed to maintain optical stability. We are currently testing two design configurations, one with two spherical lenses and one aspherical lens, and another with two aspherical lenses and one spherical lens. The ADC is designed to correct atmospheric dispersion for zenith angles ranging from 0 degree to 60 degree. The system is optimized to operate over a wavelength range of 0.4 $\mu m$ to 0.9 $\mu m$, targeting an effective field of view of about 0.5 degree. Considering the limited mechanical space available at the VBT prime focus, the design emphasizes compactness, ease of alignment, and manufacturability. The system achieves a mean D80 better than 0.3 arcsec and 0.23 arcsec for Design~1 and Design~2, respectively, at zenith, and maintains a mean D80 within 0.57 arcsec and 0.45 arcsec up to a zenith angle of 60 degree after atmospheric dispersion correction. Atmospheric dispersion at higher zenith angles (up to 60 degrees) is corrected using a movable lens element, enabling the system to preserve high image quality across the field.
\end{abstract}


\keywords{Vainu Bappu Telescope – Optical instruments – Prime focus wide-field corrector – Aspheric Surface - D80 analysis – Astronomical observing techniques }

\section{INTRODUCTION}
\label{sec:intro}  

In recent decades, wide-field imaging and spectroscopy have become essential tools in modern observational astronomy, enabling efficient surveys of large sky areas and simultaneous studies of multiple astronomical sources. The construction of newer, more advanced telescopes at sites with excellent seeing conditions has further enhanced observational capabilities worldwide. To keep older-generation telescopes competitive and scientifically relevant, it is important to improve their performance by integrating modern technologies. The Vainu Bappu Observatory (VBO) houses telescopes in the 0.5–2 m class, including the 2.34 m Vainu Bappu Telescope (VBT), which features a parabolic primary mirror and a hyperbolic secondary mirror. The prime focus, offers a relatively narrow field of view (FoV), limiting its capability for wide-field imaging and spectroscopy. However, achieving good image quality over a wide FoV is challenging because optical aberrations such as coma, astigmatism, and field curvature increase rapidly away from the optical axis. Wide Field Corrector (WFC), which consist of a system of corrective lenses placed near the prime focus, are commonly used to improve the optical performance of telescopes over a larger field.

The concept of multi-element correctors has evolved significantly over the years. Early work by Ross introduced a three-element corrector system for large telescopes (\citenum{1935ApJ....81..156R}). Later, Wynne developed an improved three-lens corrector positioned near the prime focus of parabolic telescopes to correct off-axis aberrations more effectively (\citenum{1974MNRAS.167..189W}). Since then, WFC systems have been widely adopted in many astronomical facilities to enable wide-field imaging and spectroscopy. Several telescopes employ multi-element corrector systems to achieve large corrected fields, such as the Sloan Digital Sky Survey (SDSS) telescope, the Mayall telescope, the Anglo-Australian Telescope (AAT), and the Canada-France-Hawaii Telescope (CFHT). The 2.34~m Vainu Bappu Telescope (VBT), located at the Vainu Bappu Observatory (VBO) in Kavalur, India, is an equatorially mounted reflecting telescope. Owing to its geographical location near the equator, the telescope provides access to a large fraction of both the northern and southern sky. The prime focus configuration of the VBT is particularly suitable for fiber-fed instruments and wide-field observations. The VBT is currently equipped with two spectrographs that enable spectroscopic observations of point sources over a wide range of spectral resolutions from $R \sim 2000$ to $R \sim 100000$. A fiber-fed HRES operates at the prime focus (\citenum{2005JApA...26..331R}), while the slit-based OMRS is mounted at the Cassegrain focus (\citenum{1998BASI...26..383P}). Recent large-scale sky surveys such as Gaia (\citenum{2016A&A...595A...1G}) and the Legacy Survey of Space and Time (LSST) (\citenum{2019ApJ...873..111I}) are producing extensive photometric catalogs of both point and extended astronomical sources. Spectroscopic follow-up observations of these objects are essential for investigating their physical properties and understanding the underlying astrophysical processes. The VBT supports both a fast f/3.25 beam at the prime focus. The fast prime-focus beam is particularly suitable for fiber-fed instrumentation, which enables multi-object spectroscopy and integral field spectroscopy, enabling efficient observations of multiple targets across the field. To utilize this capability, we are developing a setup to operate the HRES and OMRS spectrographs simultaneously from the prime focus (\citenum{2024SPIE13100E..4OS}).

The WFC increases the accessible FoV and allows the integration of additional back-end instruments, thereby enhancing the scientific capabilities of the telescope in the coming decade. To minimize transmission losses and satisfy mechanical constraints at the prime focus—such as limitations on weight and available space a compact optical design with a small number of elements is preferred. Accordingly, a three element WFC based on spherical lenses was previously designed and fabricated for the VBT. The detailed optical design, fabrication, and performance analysis of this system are presented in \citenum{2025ExA....60....3S}. In that study, atmospheric parameters were incorporated into the optical model to evaluate the Geometrical Encircled Energy Diameter (D80) across different zenith angles ranging from $0^{\circ}$ to $60^{\circ}$ over a $30'$ FoV. The mean geometrical D80 was found to be $1.22''$ at zenith ($0^{\circ}$), increasing to $1.40''$ at $30^{\circ}$, $1.81''$ at $50^{\circ}$, and $2.25''$ at $60^{\circ}$. The corresponding FWHM values were $0.80''$, $0.92''$, $1.18''$, and $1.47''$, respectively. These results indicate that the use of purely spherical lenses leads to relatively large spot sizes and D80 values, which can result in fiber coupling losses and image degradation. To address these limitations, In the present work, we extend that study by investigating improved optical configurations that incorporate aspherical surfaces and atmospheric dispersion correction capability. Two different three-lens configurations are explored, combining spherical and aspherical lenses to improve image quality across the field while maintaining a compact and mechanically feasible design. The optical systems are optimized to correct aberrations over a FoV of 30$'$ in the wavelength range of 0.4--0.9~$\mu$m. The paper presents the optical design methodology, the configuration of the proposed WFC systems, and a detailed analysis of their performance at different zenith angles. The results demonstrate that the use of aspherical surfaces can further improve the imaging performance of the VBT prime focus while maintaining a simple and compact optical design.

\section{Optical design of wide-field corrector system}
\label{sec:current_coupl}

The WFC is designed for the prime focus of the VBT to increase the FoV from $4'$ to $30'$. The telescope specifications and site atmospheric parameters used in this work are adopted from the previous spherical WFC design presented in \citenum{2025ExA....60....3S} and are summarized in Table~\ref{tab:vbt_combined}. The seeing conditions at the VBT site typically vary between $1.5''$ and $3.5''$, with an average value of approximately $2.5''$ (\citenum{2009MNRAS.395..593P}). Two new optical configurations have been developed and optimized using ZEMAX with different optical glass materials. In addition, aspheric surfaces are introduced to further reduce optical aberrations. At least one lens is retained as spherical to allow mechanical compensation of atmospheric dispersion through controlled decentering and tilting. The optimization process involved multiple iterations, focusing on key parameters such as the radius of curvature (RoC), center thickness (CT), and inter-lens spacing to minimize aberrations and achieve the desired image quality. The final design consists of a compact three-lens system capable of delivering improved performance over a $30'$ FoV.

Atmospheric conditions, specifically pressure, temperature, and humidity, directly influence the refractive index of air, thereby affecting atmospheric dispersion and the overall image quality of the WFC. Among these parameters, local pressure has the most significant impact on D80 performance, as it is intrinsically linked to the column density of the overlying atmosphere. Under hydrostatic equilibrium, governed by the relation (\cite{1977atsc.book.....W}),
\begin{equation}
\frac{dP}{dz} = -\rho g,
\end{equation}

where $P$ is the pressure, $\rho$ is the air density, and $g$ is the acceleration due to gravity. Integrating this equation from altitude $z$ to the top of the atmosphere gives
\begin{equation}
P(z) = \int_{z}^{\infty} \rho(z)\, g \, dz.
\end{equation}

The pressure at a given altitude is determined by the weight of the air column above it. Consequently, pressure is directly proportional to the total integrated mass (column density) of the atmosphere. Therefore, an increase in local pressure corresponds to a higher column density that the incoming starlight must traverse, leading to a more pronounced effect on image quality.

\begin{table}[htbp]
    \centering
    \caption{Prime Focus and Atmospheric Specifications of the 2.34m VBT}
    \label{tab:vbt_combined}

    \begin{subtable}[t]{0.45\textwidth}
        \centering
        \caption{VBT Prime Focus Specifications}
        \begin{tabular}{@{}l c@{}}
            \toprule
            \textbf{Parameter} & \textbf{Value} \\ 
            \midrule
            Primary Mirror Diameter & 2340~mm \\ 
            Focal Length & 7605~mm \\ 
            F/ratio & 3.25 \\ 
            Field of View (FoV) & $\approx$ 4$'$ \\ 
            Conic Constant & -1 \\ 
            Central Obscuration & 662~mm \\ 
            Image Scale & 27~arcsec/mm \\ 
            \bottomrule
        \end{tabular}
        \label{tab:vbt_spec}
    \end{subtable}
    \hfill
    \begin{subtable}[t]{0.5\textwidth}
        \centering
        \caption{VBO Site Atmospheric Parameters}
        \begin{tabular}{@{}l c@{}}
            \toprule
            \textbf{Atmospheric Parameters} & \textbf{Average Measurement} \\ 
            \midrule
            Air Temperature (K) & 296 \\ 
            Pressure (mbar) & 998.4 \\ 
            Humidity & 0.835 \\  
            Average Wind Speed (m/s) & 0.21 \\
            Altitude (m) & 725 \\ 
            Longitude & 78$^\circ49.6'$ E \\
            Latitude & 12$^\circ34.6'$ N \\
            \bottomrule
        \end{tabular}
        \label{tab:vbt_atm}
    \end{subtable}
\end{table}

\subsection{Design 1: WFC with One Aspheric and Two Spherical Lenses}

\begin{figure}[htbp]
\centering
\includegraphics[width=0.7\linewidth]{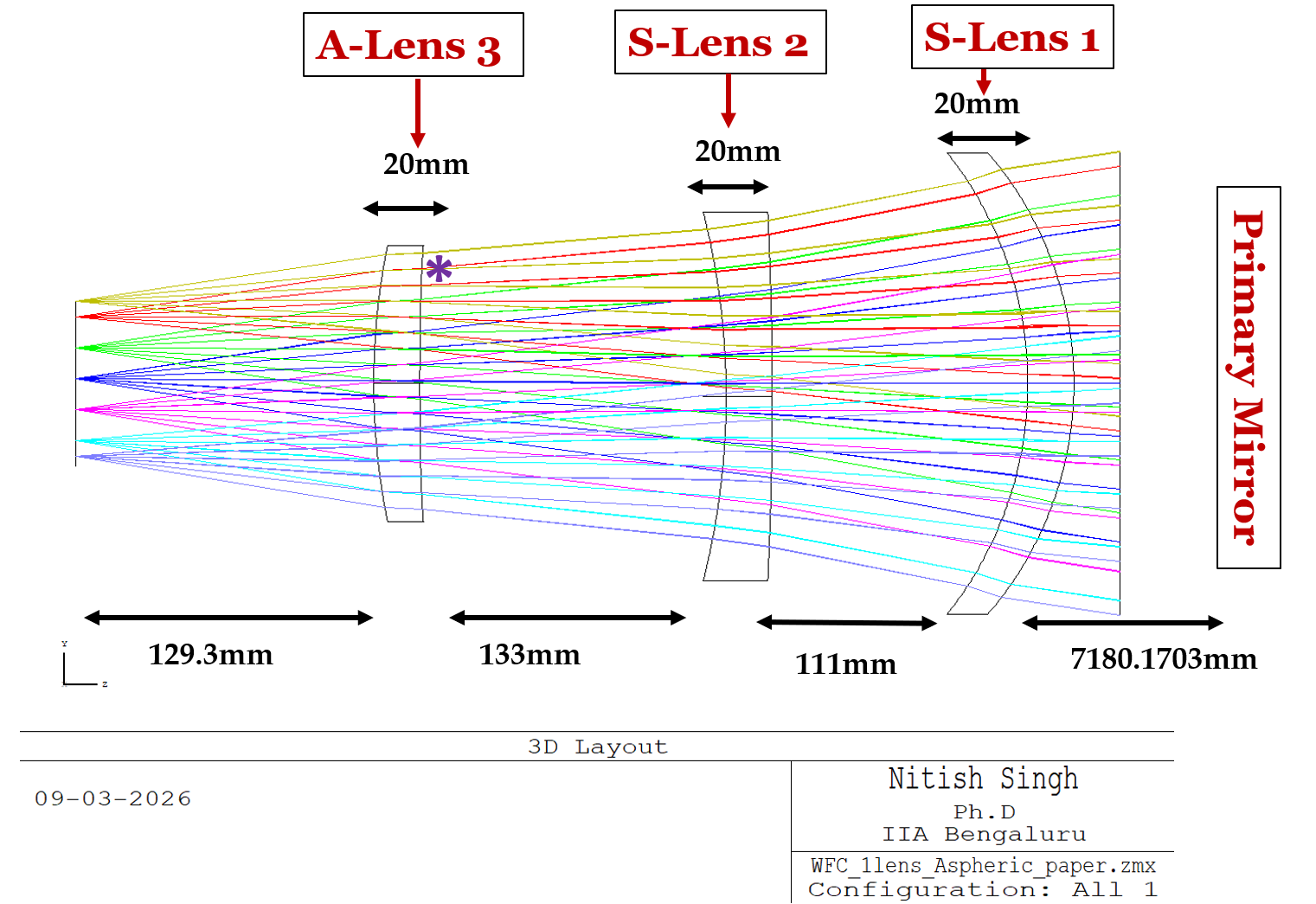}
\caption{Design 1: WFC optical layout with one aspheric and two spherical lenses. The asterisk (*) indicates the aspheric surface.}
\label{fig:1aspheric_optical_layout}
\end{figure}

The first configuration consists of a three-lens system with one aspheric lens (A-Lens~3) and two spherical lenses (S-Lens~1 and S-Lens~2), as depicted in Fig.~\ref{fig:1aspheric_optical_layout}. In this design, S-Lens~2 is the movable element and is responsible for providing the required tip, tilt, and decenter adjustments to perform ADC. Table~\ref{tab:aspheric_parameters} summarizes the corresponding WFC unit movements and the Y-decenter values needed for S-Lens~2 at different zenith angles.

\subsection{Design 2: WFC with Two Aspheric and One Spherical Lens}

\begin{figure}[htbp]
\centering
\includegraphics[width=0.7\linewidth]{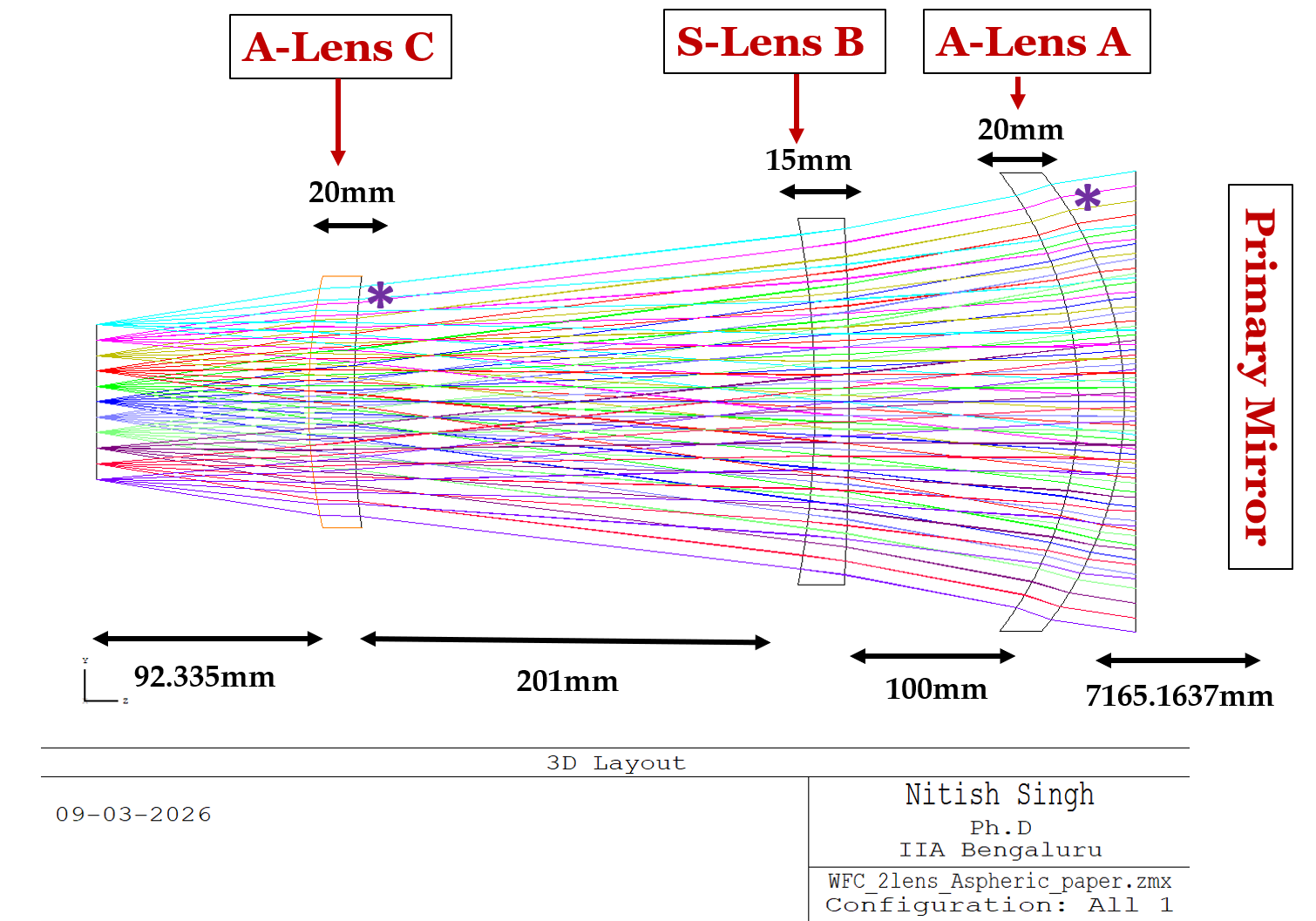}
\caption{Design 2: WFC optical layout with two aspheric and one spherical lens. The asterisk (*) indicates the aspheric surface.}
\label{fig:2aspheric_optical_layout}
\end{figure}

The second configuration uses a three-lens system with two aspheric lenses (A-Lens~A and A-Lens~C) and one spherical lens (S-Lens~B), as shown in Fig.~\ref{fig:2aspheric_optical_layout}. Here, S-Lens~B serves as the movable element, providing the tip, tilt, and decenter adjustments necessary for ADC. The required WFC unit movements and Y-decenter values for S-Lens~B at different zenith angles are listed in Table~\ref{tab:aspheric_parameters}.

 Both design lenses have diameters less than 200~mm, as shown in Figs.~\ref{fig:1aspheric_optical_layout} and \ref{fig:2aspheric_optical_layout}. The optimized optical parameters, including RoC of convex surface (CVS) and RoC of concave surface (CCV), CT, clear aperture diameter (CA Dia), edge thickness (ET), and material thickness (MT), are summarized in Tables~\ref{tab:wfc_design_parameters}. The corresponding movements required at different zenith angles for this configuration are summarized in Table~\ref{tab:aspheric_parameters}.

\begin{table}[htbp]
\centering
\caption{Required movement of the WFC unit and Y-decenter for different zenith angles for two configurations: Design~1 (two spherical lenses) and Design~2 (two aspherical lenses).}
\begin{tabular*}{\textwidth}{@{\extracolsep\fill}lcccccc}
\toprule
& \multicolumn{2}{c}{\textbf{Design 1 }} 
& \multicolumn{2}{c}{\textbf{Design 2}} \\
\cmidrule(lr){2-3} \cmidrule(lr){4-5}
\textbf{Parameter} & \textbf{Full WFC Unit} & \textbf{Y-Decenter} 
& \textbf{Full WFC Unit} & \textbf{Y-Decenter}  \\
\midrule
0$^\circ$ & 0 & 0 & 0 & 0 \\
30$^\circ$ & -0.0035189 & -1.92909 & -0.0043012 & -2.77339  \\
50$^\circ$ & -0.01343 & -3.96975 & -0.01742 & -5.62445  \\
60$^\circ$ & -0.02691 & -5.72091 & -0.03542 & -8.03601  \\

\bottomrule
\end{tabular*}
\label{tab:aspheric_parameters}
\end{table}

The detailed optical design parameters for both configurations are summarized in Table~\ref{tab:wfc_design_parameters}. The asterisk (*) denotes the aspheric surface in the WFC design. In our previous work, we successfully designed and fabricated lenses with high curvature using spherical surfaces. The measured surface figure error (SFE) was better than $0.05\,\lambda$, demonstrating good fabrication quality. In the present work, the introduction of aspheric surfaces adds an additional level of complexity. Maintaining the required asphericity along with precise curvature control is one of the primary challenges in the fabrication and alignment process. 

\begin{table}[!ht]
\centering
\caption{Optical design parameters of the WFC lenses for two configurations: Design~1 (two spherical lenses) and Design~2 (two aspherical lenses).}
\begin{tabular*}{\textwidth}{@{\extracolsep\fill}lcccccc}
\toprule
& \multicolumn{3}{c}{\textbf{Design 1}} 
& \multicolumn{3}{c}{\textbf{Design 2}} \\
\cmidrule(lr){2-4} \cmidrule(lr){5-7}
\textbf{Parameter} & \textbf{S-Lens 1} & \textbf{S-Lens 2} & \textbf{A-Lens 3} 
& \textbf{A-Lens A} & \textbf{S-Lens B} & \textbf{A-Lens C} \\
\midrule
CA Dia (mm) & 200 & 160 & 120 & 200 & 160 & 120 \\
RoC of CVS (mm) & 152.966 & 1669.848 & 304.329 & 157.503 (*) & 1437.596 & 246.753 \\
RoC of CCS (mm) & 160.850 & 329.044 & 2331.861 (*) & 163.099 & 442.883 & 617.658 (*) \\
CT (mm) & 20 & 20 & 20 & 20 & 15 & 20 \\
ET (mm) & 17.608 & 27.79 & 15.514 & 18.37 & 21.34 & 17.03 \\
MT (mm) & 53.91 & 29.39 & 22.68 & 54.17 & 24.25 & 23.36 \\
RMS Concave SFE & 0.05~$\lambda$  & 0.05~$\lambda$ & 0.05~$\lambda$ & 0.05~$\lambda$  & 0.05~$\lambda$ & 0.05~$\lambda$   \\
RMS Convex SFE & 0.05~$\lambda$   &  0.05~$\lambda$ & 0.05~$\lambda$ & 0.05~$\lambda$   &  0.05~$\lambda$ & 0.05~$\lambda$    \\
RMS Wavefront Error & 0.07~$\lambda$ & 0.07~$\lambda$  & 0.07~$\lambda$  & 0.07~$\lambda$ & 0.07~$\lambda$  & 0.07~$\lambda$\\ 

Glass Material & K3 & SSK1 & PK2 & K3 & SSK1 & PK2 \\
\bottomrule
\end{tabular*}
\label{tab:wfc_design_parameters}
\end{table}

The aspheric surfaces used in the optical designs are described by higher-order polynomial coefficients. The corresponding aspheric coefficients are given below.

A-Lens A (*): 4th-order coefficient = $-1.30767 \times 10^{-10}$ and 
6th-order coefficient = $7.15691 \times 10^{-15}$.

A-Lens B (*): 4th-order coefficient = $9.62454 \times 10^{-8}$ and 
6th-order coefficient = $6.09555 \times 10^{-13}$.

A-Lens 3 (*): 4th-order coefficient = $5.56502 \times 10^{-8}$ and 
6th-order coefficient = $9.66839 \times 10^{-13}$.

\section{Opto-Mechanical Tolerance Analysis For Design 1 and Design 2}

The tolerance analysis for the present designs has been carried out by considering both optical sensitivity and our established fabrication capabilities from the previously developed spherical WFC. The tolerance requirements for WFC Design~1 and Design~2 are summarized in Table~\ref{tab:tolerance_combined}. The adopted tolerances include manufacturing limits on parameters such as the variations in the RoC of both convex and concave surfaces, CT, surface decenter, and surface irregularity Peak-to-Valley (PV).
In our previous work, we successfully fabricated and tested spherical lenses for a three-element WFC, achieving a surface figure error better than $0.05\,\lambda$, demonstrating high manufacturing accuracy. Therefore, the tolerance values adopted for the spherical surfaces in the present designs are based on the demonstrated fabrication performance reported in our earlier work (\citenum{2025ExA....60....3S}).

An aspheric surface accuracy tolerance of $\pm 50~\mu$m has been assigned, representing the maximum allowable deviation of the fabricated surface from the designed aspheric profile. In addition, alignment tolerances were defined in terms of allowable decenter along the X, Y, and Z directions, as well as tilt about the X and Y axes. These tolerances were determined based on sensitivity analysis while ensuring practical feasibility in fabrication and assembly, guided by our prior experience with spherical lens systems. A compensator of $\pm10$~mm has been assigned to the Back Focal Length (BFL) to accommodate cumulative deviations arising from manufacturing and alignment errors. This tolerance framework ensures that the optical performance of the system remains within acceptable limits during fabrication and integration, without introducing unnecessary constraints on manufacturability.
\begin{table}[htbp]
\centering
\caption{Manufacturing and alignment tolerances for the WFC designs.}
\begin{tabular*}{\textwidth}{@{\extracolsep\fill}lcc}
\toprule
\textbf{Tolerance Parameter} & \textbf{Design 1} & \textbf{Design 2} \\
\hline

RoC (CCS/CVS) (mm) & $\pm 0.5$ & $\pm 0.5$ \\
Centre Thickness (mm) & $\pm 2$ & $\pm 2$ \\
Surface Decenter (arcmin) & $<1$ & $<1$ \\
Surface Irregularity (PV) & $<0.2\,\lambda$ & $<0.2\,\lambda$ \\

Decenter in X \& Y ($\mu$m) & $\pm 100$ & $\pm 100$ \\
Decenter in Z ($\mu$m) & $\pm 50$ & $\pm 50$ \\
Tilt (arcsec) & $\pm 60$ & $\pm 60$ \\

BFL Compensation (mm) & $\pm 10$ & $\pm 10$ \\

\bottomrule
\end{tabular*}
\label{tab:tolerance_combined}
\end{table}

\section{Performance Analysis}

To evaluate the optical performance of the proposed WFC designs (design 1 and design 2 as shown in Fig.~\ref{fig:1aspheric_optical_layout}, \ref{fig:2aspheric_optical_layout}) and atmospheric parameters listed in Table~\ref{tab:vbt_atm} were incorporated into the optical model. The performance was analyzed in terms of the geometrical Encircled Energy Diameter (D80) across a 30$'$ FoV for zenith angles ranging from 0$^{\circ}$ to 60$^{\circ}$. The primary optimization goal during this analysis was to achieve a minimum, nearly ideal geometrical spot size across the entire field, ensuring that the WFC optics introduce negligible degradation to the incoming wavefront. The Full Width at Half Maximum (FWHM) was estimated using the relation given in \citenum{2019sto..book.....T}:

\begin{equation}
\text{D80 = FWHM} \times 1.524
\label{eq:net_seeing}
\end{equation}

The performance of both optical configurations is discussed below.

\subsection{Performance Analysis for Design 1}

\begin{figure}[htbp]
\centering
\includegraphics[width=1\linewidth]{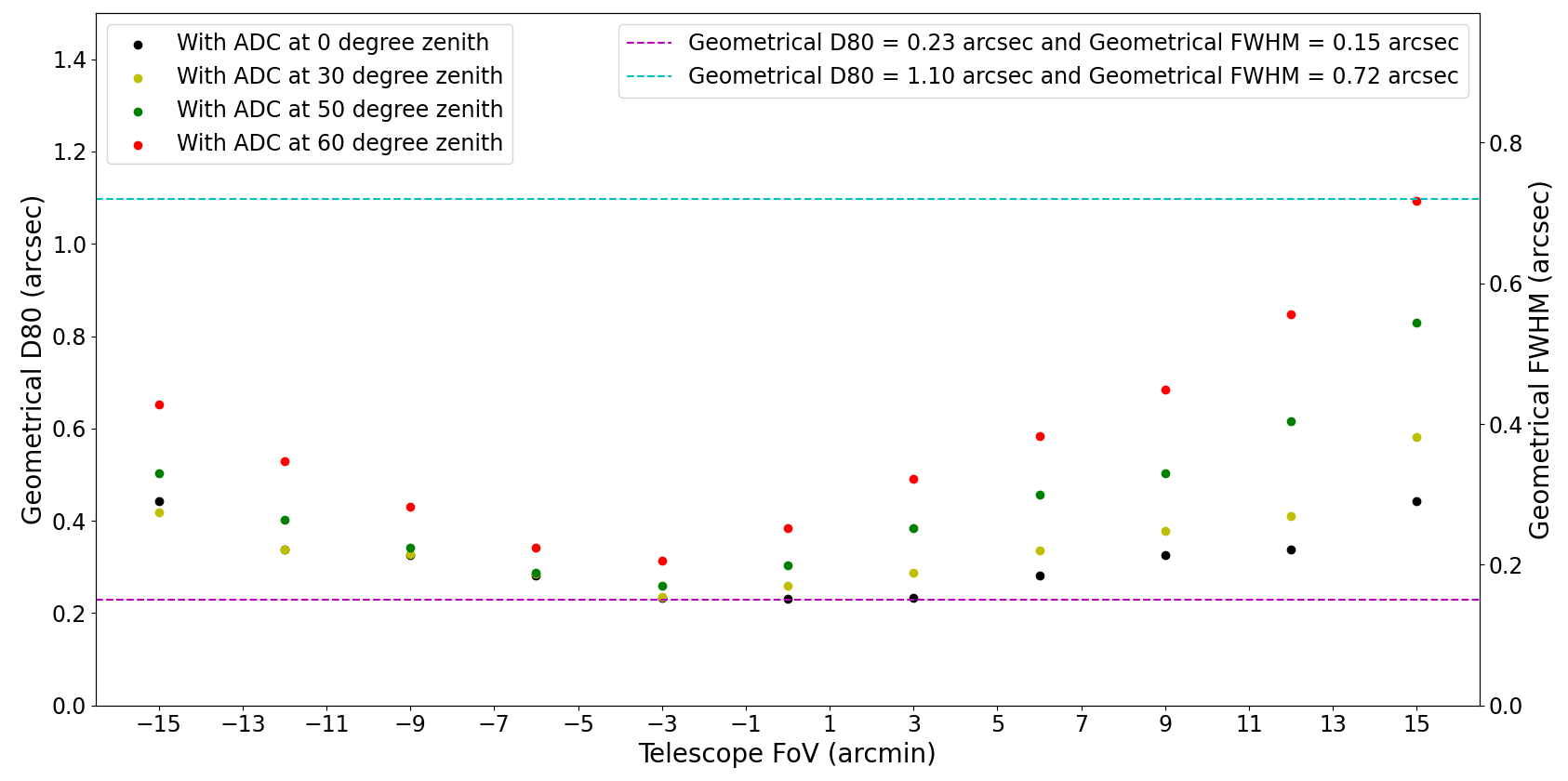}
\caption{Design 1: Geometrical D80 performance across a 30$'$ FoV for zenith angles from $0^{\circ}$ to $60^{\circ}$ over a wavelength range of 0.4--0.9~$\mu$m for.}
\label{fig:design1_spot}
\end{figure}

After introducing Design~1, the effective focal ratio becomes f/3.26 with a corresponding plate scale of $37.11~\mu$m/arcsec. At zenith ($0^\circ$), the WFC demonstrates excellent image quality across the full FoV, with a mean geometrical D80 of $0.31''$ and a corresponding mean FWHM of $0.20''$. The performance of the ADC is evaluated at higher zenith angles. The mean D80 increases to $0.35''$ at $30^\circ$, $0.44''$ at $50^\circ$, and $0.57''$ at $60^\circ$. Similarly, the mean FWHM increases to $0.22''$ at $30^\circ$, $0.29''$ at $50^\circ$, and $0.37''$ at $60^\circ$. Despite the increasing zenith angle, the ADC effectively compensates for atmospheric dispersion, maintaining good image quality across the field. The variation of D80 and FWHM across the field for different zenith angles is shown in Fig.~\ref{fig:design1_spot}.

\subsection{Performance Analysis for Design 2}

\begin{figure}[htbp]
\centering
\includegraphics[width=1\linewidth]{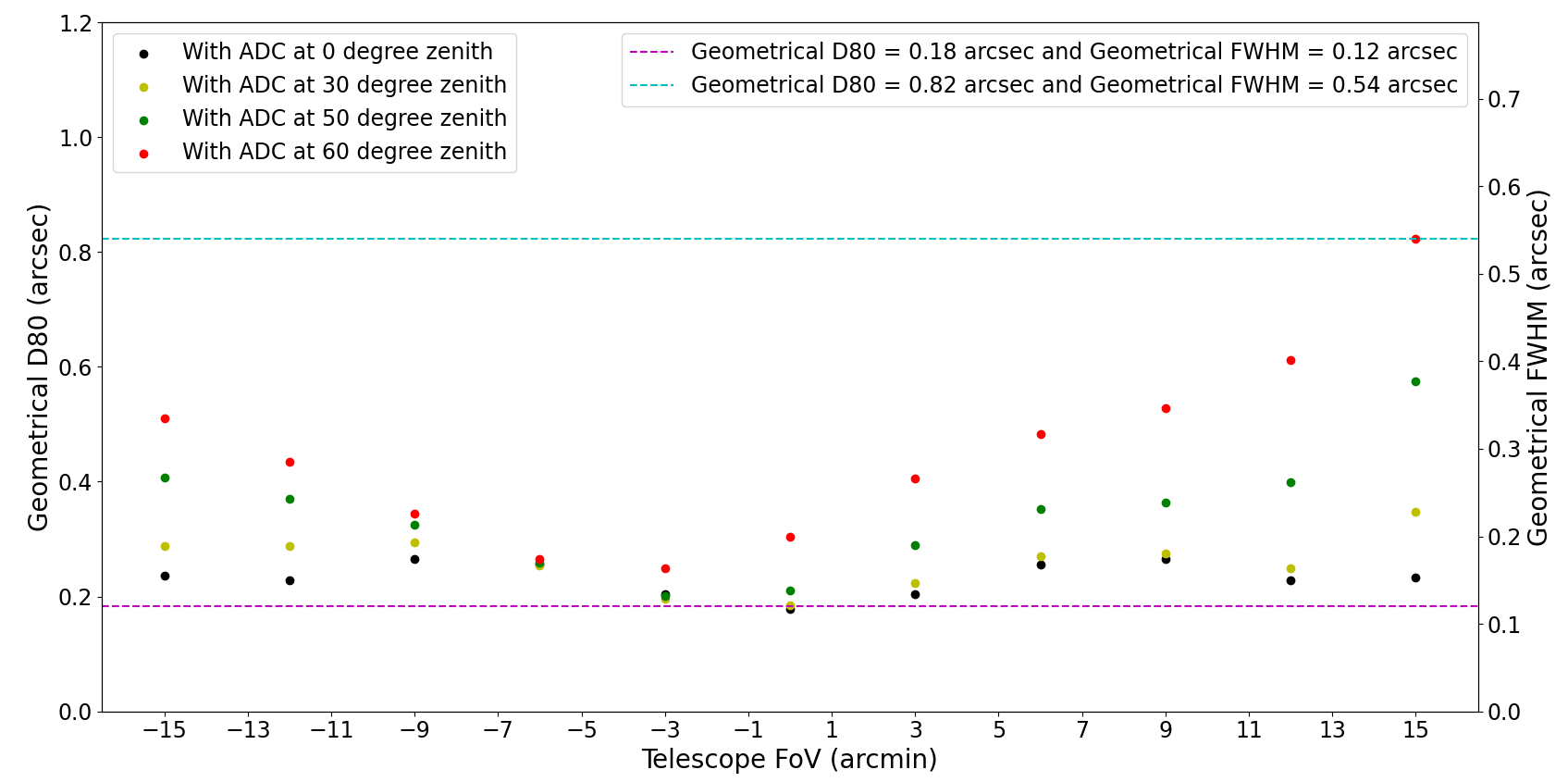}
\caption{Design 2: Geometrical D80 performance across a 30$'$ FoV for zenith angles from $0^{\circ}$ to $60^{\circ}$ over a wavelength range of 0.4--0.9~$\mu$m.}
\label{fig:design2_spot}
\end{figure}

After introducing Design~2, the effective focal ratio becomes f/3.27, with a corresponding plate scale of $37.22~\mu$m/arcsec. At zenith ($0^\circ$), the WFC exhibits excellent image quality across the full FoV, with a mean geometrical D80 of $0.23''$ and a corresponding mean FWHM of $0.15''$. The performance of the ADC is evaluated at higher zenith angles. The mean D80 increases to $0.26''$ at $30^\circ$, $0.34''$ at $50^\circ$, and $0.45''$ at $60^\circ$. Similarly, the mean FWHM increases to $0.17''$ at $30^\circ$, $0.22''$ at $50^\circ$, and $0.29''$ at $60^\circ$. Despite the increasing zenith angle, the ADC effectively compensates for atmospheric dispersion, maintaining high image quality across the field. The variation of D80 and FWHM across the field for different zenith angles is shown in Fig.~\ref{fig:design2_spot}.

After introducing atmospheric parameters from Table~\ref{tab:vbt_atm}, both Design~1 and Design~2 show symmetric D80 performance at zenith ($0^\circ$) over the field range of $-15'$ to $+15'$. However, at higher zenith angles (30$^\circ$ and 60$^\circ$), the D80 performance becomes asymmetric, with the $+15'$ field performing worse than the $-15'$ field (see Fig.~\ref{fig:design1_spot} and \ref{fig:design2_spot}). This asymmetry arises due to atmospheric refraction. As light travels from vacuum ($n_v = 1$) into Earth's atmosphere ($n_a \approx 1.0003$), it bends according to Snell’s law \cite{1999prop.book.....B}:

\begin{equation}
    n_v \sin(\theta_v) = n_a \sin(\theta_a)
\end{equation}

At a zenith angle of 60$^\circ$ with a FoV of $\pm 15'$, the effective incident angles differ slightly (59.75$^\circ$ and 60.25$^\circ$). Due to the nonlinear nature of refraction, this results in unequal bending, with the $+15'$ ray refracting more than the $-15'$ ray. This differential refraction leads to the observed asymmetry in D80 performance.

We also observed that variation in the WFC performance across different elevation angles is fundamentally driven by changes in the optical path length of light traversing the Earth's atmosphere. In observational astronomy, this path length is quantified as Air Mass ($X$). At zenith (0$^\circ$), the telescope points directly upward, and incident light passes through the thinnest possible cross section of the atmosphere ($X = 1$). This minimal optical path yields the lowest levels of atmospheric scattering and chromatic dispersion, resulting in optimal image quality and the tightest geometrical D80 spot sizes. As the zenith angle ($z$) increases, the observation path becomes increasingly slanted. For instance, at a zenith angle of 60$^\circ$, incident light must travel through roughly twice as much atmospheric volume to reach the primary mirror. For moderate angles, this relationship is well-approximated by a simple secant function:

$$X \approx \frac{1}{\cos(z)}$$

Consequently, at 60$^\circ$, the Air Mass effectively doubles ($X \approx 2$). This extended transit through the atmospheric column proportionally exacerbates atmospheric dispersion, the wavelength dependent refraction of light, which geometrically smears the stellar profile and inflates the D80 spot size at the focal plane.

\section{Transmission Efficiency of the design WFC}

Both Design 1 and Design 2 employ a three element corrector, consisting of a combination of spherical and aspheric lenses. Since the transmission efficiency of an optical element is primarily governed by the material properties and surface reflections, as described by the Fresnel equations, it is largely independent of the surface geometry (spherical or aspheric) (\citenum{1999prop.book.....B, 2017opti.book.....H}). Therefore, the transmission efficiency of the WFC system was evaluated over the wavelength range of 0.4~$\mu$m to 0.9~$\mu$m. The WFC consists of three lenses, resulting in a total of six air glass interfaces. The theoretical transmission of the WFC was estimated using the Sellmeier equations for each optical element using ZEMAX (\citenum{Zemax}). For a glass element, the refractive index $n(\lambda)$ as a function of wavelength $\lambda$ is given by the Sellmeier formula:

\begin{equation}
n^2(\lambda) = 1 + \frac{B_1 \lambda^2}{\lambda^2 - C_1} + \frac{B_2 \lambda^2}{\lambda^2 - C_2} + \frac{B_3 \lambda^2}{\lambda^2 - C_3},
\end{equation}

where $B_1$, $B_2$, $B_3$ and $C_1$, $C_2$, $C_3$ are the material-specific Sellmeier coefficients, and $\lambda$ is in microns. The reflection at each air-glass interface is approximated using the Fresnel equation at normal incidence (\citenum{1999prop.book.....B, 2017opti.book.....H}):

\begin{equation}
R(\lambda) = \left( \frac{n(\lambda) - 1}{n(\lambda) + 1} \right)^2,
\end{equation}

and the corresponding transmission of entire Lens (Two Interfaces) is $T(\lambda)$, assuming no absorption within the bulk material.

\begin{equation}
T(\lambda) = \left[ 1 - R(\lambda) \right]^2,
\end{equation}

For our WFC, composed of three optical elements SSK1, PK2, and K3, the total transmission without any anti-reflection coating is then

\begin{equation}
T_\mathrm{total}(\lambda) = T_\mathrm{SSK1}(\lambda) \cdot T_\mathrm{PK2}(\lambda) \cdot T_\mathrm{K3}(\lambda),
\end{equation}

where $T_\mathrm{SSK1}$, $T_\mathrm{PK2}$, and $T_\mathrm{K3}$ are the transmissions of the respective glass elements. The refractive indices of the WFC optical elements were computed using the Sellmeier equation, with coefficients obtained from the Schott catalog as implemented in Zemax (\citenum{Zemax}). Table~\ref{tab:sellmeier_coeff} lists the coefficients for SSK1, PK2, and K3 glasses.

\begin{table}[htbp]
\centering
\caption{Sellmeier coefficients ($B_1$, $B_2$, $B_3$, $C_1$, $C_2$, $C_3$) for the glasses used in the WFC. Values are taken from the Schott catalog (via Zemax).}
\label{tab:sellmeier_coeff}
\begin{tabular}{lcccccc}
\hline
Glass & $B_1$ & $B_2$ & $B_3$ & $C_1$ & $C_2$ & $C_3$ \\
\hline
SSK1 & 1.42691045 & 0.141195647 & 0.986004821 & 0.00831214602 & 0.0334369911 & 109.701359 \\
PK2  & 1.20921993 & 0.0668873331 & 1.01818125 & 0.00733975047 & 0.0286495917 & 101.852918 \\
K3   & 1.08999358 & 0.181850015 & 0.936478693 & 0.00670233559 & 0.0254232565 & 113.486321 \\
\hline
\end{tabular}
\end{table}

\noindent These coefficients were used in conjunction with the Sellmeier equation to calculate the wavelength-dependent refractive index $n(\lambda)$, which was then used to estimate the Fresnel reflection and total transmission of the uncoated WFC.

\begin{figure}[htbp]
\centering
\includegraphics[width=0.9\linewidth]{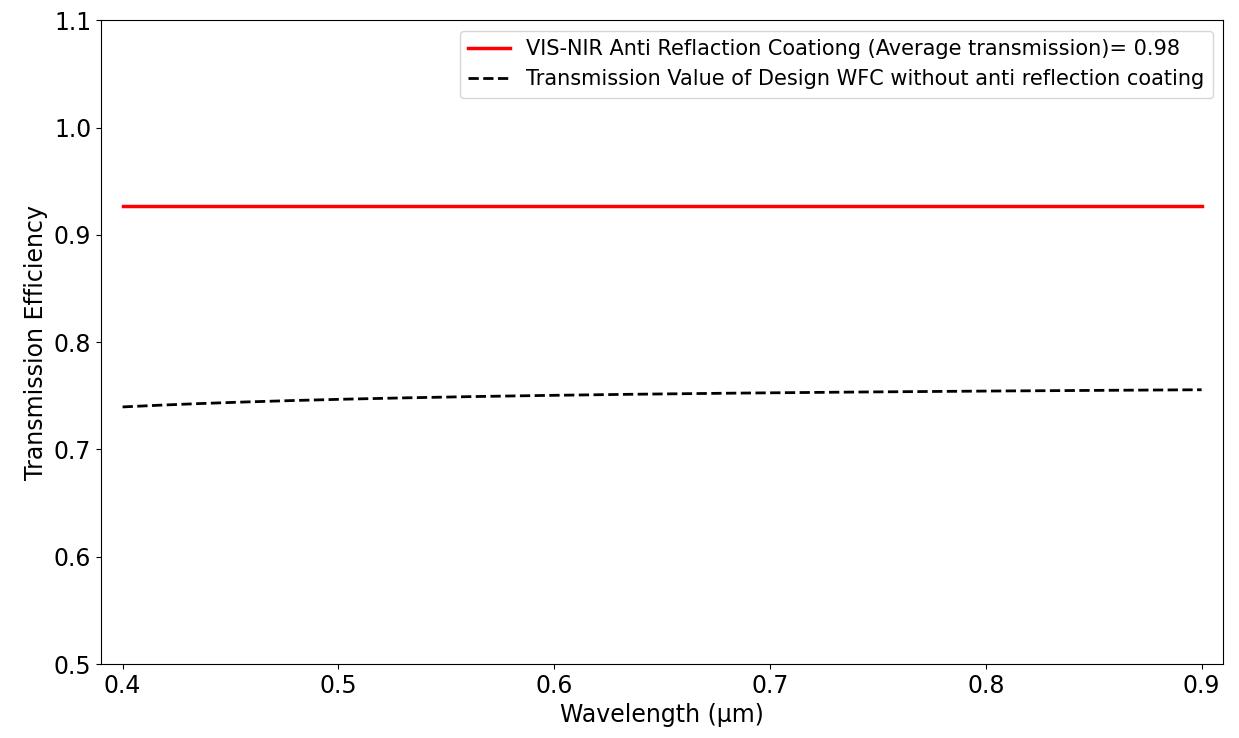}
\caption{Theoretical transmission efficiency of the uncoated three-element WFC as a function of wavelength.}
\label{fig:transmission_aspheric}
\end{figure}

The estimated transmission of the uncoated WFC is approximately, starting from approximately $73.97\%$ at wavelength $0.4~\mu m$, rising to $\approx 74.89\%$ at wavelength $0.55~\mu m$, reaching $\approx 75.37\%$ at wavelength $0.75~\mu m$, and finally attaining $\approx 75.57\%$ at wavelength $0.9~\mu m$ (see Fig.~\ref{fig:transmission_aspheric}). To further enhance transmission efficiency, we plan to apply a broadband anti-reflection coating. The \textit{VIS-NIR coating from Edmund Optics} \footnote{\url{https://www.edmundoptics.in/knowledge-center/application-notes/lasers/anti-reflection-coatings/?srsltid=AfmBOorZN9cdFAvqJLOQyN_38fGu1ipANpjY7KC_hZD_PfhEqZVW44Yj}}, with specifications of $T_{\text{avg}} \leq 98.75\%$ ($R_{\text{avg}} \leq 1.25\%$)  across 0.4–1.0~$\mu m$ for each lens and over all WFC transmission will be $\approx$92.73\% , is expected to improve the overall transmission performance.

\section{Conclusion}

We have designed a compact and lightweight three-element WFC system using both spherical and aspherical lenses for the VBT prime focus. The system, composed of three optical lenses (spherical and aspherical) covering the polychromatic wavelength range 0.4--0.9~$\mu$m, enhances the corrected FoV from 4$'$ to 30$'$ at the VBT prime focus. Aberration correction, tolerance analysis, and atmospheric dispersion were performed using ZEMAX to ensure precision and stability within defined limits. Manufacturing and alignment tolerances were also analyzed, as summarized in Table~\ref{tab:wfc_design_parameters}. The transmission efficiency of the WFC was calculated using the Sellmeier equations, yielding approximately $73.97\%$ at 0.4~$\mu$m, $74.89\%$ at 0.55~$\mu$m, $75.37\%$ at 0.75~$\mu$m, and $75.57\%$ at 0.9~$\mu$m (see Fig.~\ref{fig:transmission_aspheric}). With a VIS-NIR anti-reflection coating applied to the lenses, the overall transmission is expected to increase to $\sim 92\%$. The D80 performance provides a key measure of image quality for both designs. For Design~1, the mean geometrical D80 obtained from ZEMAX simulations is 0.31$''$ (11.50~$\mu$m) at zenith (0$^\circ$), increasing to 0.35$''$ at 30$^\circ$ zenith angle and 0.57$''$ at 60$^\circ$ (see Fig.~\ref{fig:design1_spot}). For Design~2, the mean geometrical D80 is 0.23$''$ (8.56~$\mu$m) at zenith, 0.26$''$ at 30$^\circ$, and 0.45$''$ at 60$^\circ$ (see Fig.~\ref{fig:design2_spot}).  

We compared our WFC (design 1 and design 2) with similar systems on other telescopes. The 2.2~m UHWFI achieves a D80 of 0.69$''$ over 0.4–1.0~$\mu$m \cite{2006PASP..118..780H}, the 4.01~m Mayall Telescope provides a 1.5$^\circ$ corrected FoV with D80 of 0.5$''$ at zenith and 1.5$''$ at 60$^\circ$ \cite{2024AJ....168...95M, 2014SPIE.9151E..1MS}, and the 3.9~m AAT achieves D80 of 0.5$''$ at zenith and 1.5$''$ at 60$^\circ$ \cite{1994ApOpt..33.7362J}. The CFHT demonstrates a D80 of 0.3$''$ at zenith over 0.35–2.0~$\mu$m. Also the 8.1~m Gemini achieves 0.49$''$ at zenith and 0.62$''$ at 70$^\circ$ \cite{2004SPIE.5492..841L}. Our previously fabricated spherical-lens WFC for the 2.34~m VBT prime focus achieved a 0.5$^\circ$ FoV with D80 of 1.22$''$ at zenith and 2.25$''$ at 60$^\circ$ \cite{2025ExA....60....3S}.  

The new designs, incorporating aspherical lenses and an ADC, demonstrate significantly improved performance. Design~2 exhibits the highest overall optical fidelity, while Design~1 remains a highly competitive alternative with slightly reduced manufacturing complexity. Overall, aspherical lens systems perform better than purely spherical designs. Overall, the proposed WFC provides an efficient and practical solution for enabling wide-field imaging and multi-object spectroscopic capabilities at the VBT. Design~2 offers the best optical performance, while Design~1 remains a viable alternative with reduced fabrication complexity, making both configurations valuable depending on implementation constraints.

\section{Acknowledgment}

This research was supported by the Tata Consultancy Services (TCS) Research Fellowship and the Indian Institute of Astrophysics (IIA), under the Department of Science and Technology (DST), Government of India.

\bibliography{report} 
\bibliographystyle{spiebib} 

\end{document}